\documentclass[twocolumn]{emulateapj}
\submitted{Science with Parkes @ 50 Years Young, 31 Oct. -- 4 Nov., 2011}

\usepackage{bibentry,natbib}
\shorttitle{Pulsars at Parkes}
\shortauthors{R. N. Manchester}

\begin{document}

\title{Pulsars at Parkes}

\author{R. N. Manchester}

\affil{CSIRO Astronomy and Space Science, Australia
  Telescope National Facility, P.O. Box 76, Epping, NSW 1710,
  Australia} 

\begin{abstract}
  The first pulsar observations were made at Parkes on March 8, 1968,
  just 13 days after the publication of the discovery paper by Hewish
  and Bell. Since then, Parkes has become the world's most succesful
  pulsar search machine, discovering nearly two thirds of the known
  pulsars, among them many highly significant objects. It has also
  led the world in pulsar polarisation and timing studies. In this
  talk I will review the highlights of pulsar work at Parkes from
  those 1968 observations to about 2006 when the Parkes Multibeam
  Pulsar Survey was essentially completed and the Parkes Pulsar Timing
  Array project was established.
\end{abstract}

\keywords{pulsars:general --- instrumentation:miscellaneous ---
methods:observational}

\section{The early years}
The discovery of pulsars by Antony Hewish, Jocelyn Bell, et
al. \nocite{hbp+68} was announced in the 1968 February 24 issue of
Nature. Just 13 days later, on March 8, the first Parkes pulsar
observations were made by a combined CSIRO Radiophysics/Sydney
University team led by Brian Robinson and including Tom Landecker who
is present at this meeting. By great coincidence, I commenced
employment at Parkes on February 12, just 12 days before the discovery
announcement. As a raw young pre-post-doc nobody knew, I stood at
the back of the control room on March 8 and watched the first pulses
from CP1919 come through. This was the burst that ultimately appeared
on the first Australian 50-dollar note along with the telescope itself
and various other astronomy-related images. It was a lucky
scintillation maximum and I don't think the pulsar was seen as clearly
again during those observations.

\begin{figure}
\resizebox{\hsize}{!}{\includegraphics{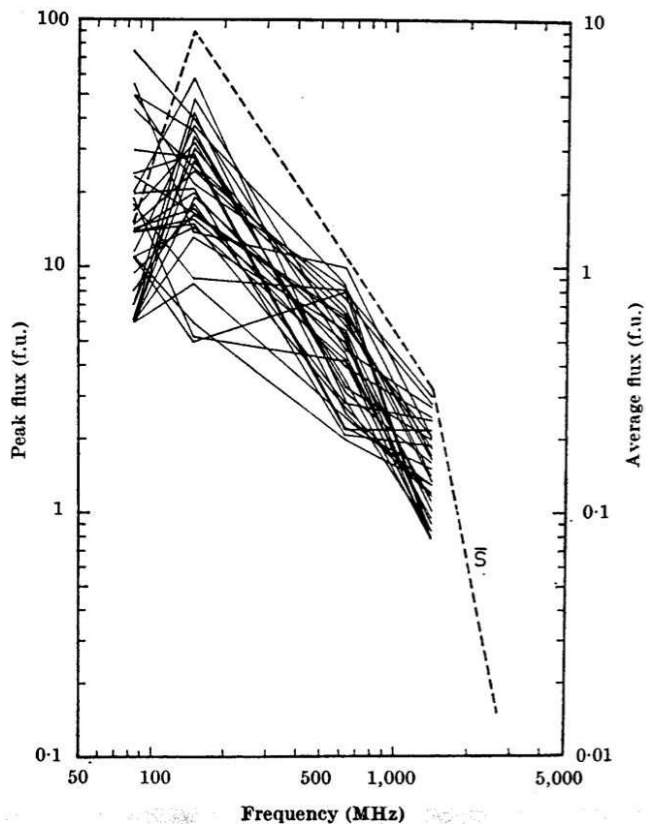}}
\caption{Observations of the spectra of individual pulses from CP1919
  (PSR B1919+21) made at Parkes on March 8, 1968, by
  \citet{rcg+68}. Individual spectra extend from 85 MHz to 1400
  MHz and show the broad-band nature of pulsar emission, the steep
  radio spectra at high frequencies and the low-frequency
  turn-over. The dashed line is the averaged spectrum which extends to
  2700 MHz and shows that the spectrum becomes increasingly steep at higher
  frequencies.  }\label{fg:cp1919}
\end{figure}

These March 8 observations \citep{rcg+68} were a ``rapid-response''
triumph -- probably the best-ever example of the versatility of a
single-dish radio telescope. Four receivers covering bands from 85 MHz
to 2700 MHz were able to observe the pulsar simultaneously. As
Fig.~\ref{fg:cp1919} shows, the quality of the resulting
individual-pulse observations was amazing and gave what is still one
of the best, if not the best, sets of such multi-frequency data.

These results were published in Nature on June 22, but this was not the
first pulsar publication from Parkes. \citet{rkc68} used the same
multi-frequency observations reported by \citet{rcg+68} to measure an
accurate period for CP1919. The Parkes telescope was able to track the
source for three hours per day over several consecutive days, whereas
the Cambridge measurements were mostly based on 4~min observations at
transit each day. The Parkes observations, published in Nature on April 20,
showed that the Cambridge group got the period of CP1919 wrong by one
pulse per day, again illustrating the benefits of a large
fully steerable single-dish radio telescope.

\begin{figure}
\resizebox{\hsize}{!}{\includegraphics{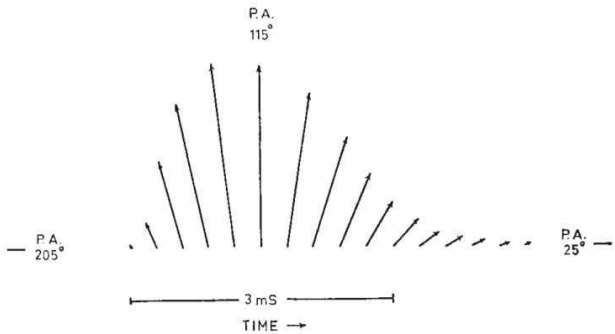}}
\caption{Observed swing of the linear polarisation vector across the Vela
  pulsar pulse profile \citep{rckm69}. The plotted vectors represent
  both the strength and angle of the linear polarisation. 
}\label{fg:vela_poln}
\end{figure}

\begin{figure}
\resizebox{\hsize}{!}{\includegraphics{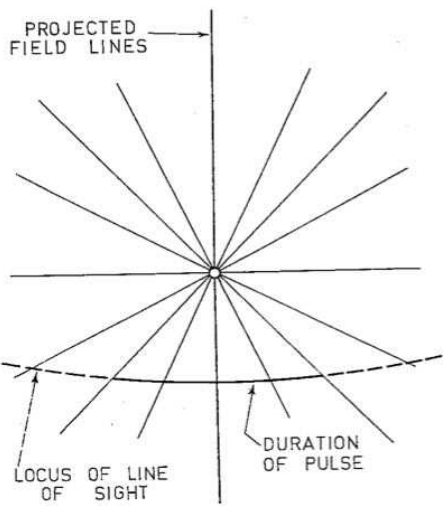}}
\caption{The rotating-vector model for pulsar emission
  \citep{rc69a}. The emission is assumed to be radiated tangentially
  to field lines in the vicinity of a magnetic pole and to have its
  polarisation position angle determined by the projected direction of
  the field lines. As the star rotates, emission is seen from regions
  along a locus which traverses the polar regions of the star. 
}\label{fg:rvm}
\end{figure}

Although other pulsar work was done at Parkes in the meantime, the
next major published result was the detection by \citet{rckm69} of the
very high linear polarisation of the Vela pulsar mean pulse profile
and the swing of polarisation position angle (PA) through the pulse
(Fig.~\ref{fg:vela_poln}). These observations strongly supported the
rotating neutron-star model where the emission beam is directed
outward from a magnetic pole and the observed PA is related to the
projected direction of the magnetic field lines \citep{rc69a}. This
idea became famous and is now known as the ``rotating-vector model''
for pulsar polarisation. As Fig.~\ref{fg:rvm} shows, this model
predicts that PA swings will be more rapid when the line of sight
passes close to the direction of the magnetic axis as the star
rotates. Subsequent observations of other pulsars provided many
examples of PA swings conforming to this model.

\begin{figure}
\resizebox{\hsize}{!}{\includegraphics{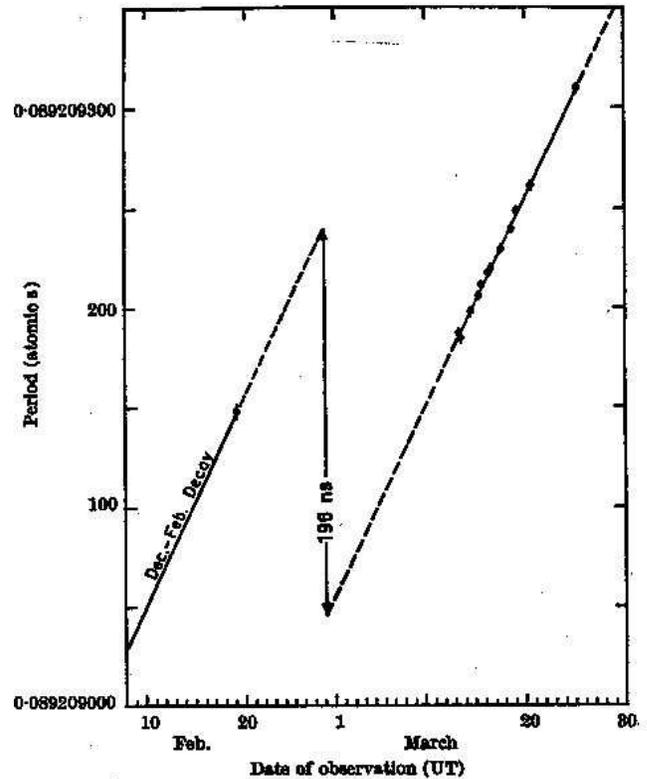}}
\caption{The first detection of a pulsar period glitch, in the Vela
  pulsar using observations made at Parkes in 1969 March. Between 1968
  December and 1969 February, the period had been steadily increasing
  at about 10.7 ns per day. Observations in March still showed this steady
  increase but, between the two sets of observations, the period had
  decreased by 196 ns or $\sim 2.3\times10^{-6}$ of the pulse period
  \citep{rm69}.}\label{fg:vela_glitch}
\end{figure}

As I had some familiarity with using the OH receiver for polarisation
observations (from my work with Brian Robinson and Miller Goss
on OH-line studies), I had been asked to help Rad with the Vela
observations. We set up the system, including the RIDL signal averager
that was used to integrate the signals synchronously with the pulsar
period, and pointed the telescope at the Vela pulsar. To our
consternation, the pulses on the RIDL were drifting across the screen,
indicating an error in the folding period for the signal
averaging. Exhaustive checks of the system and observations of other
pulsars revealed no problems and eventually we were forced to conclude
that the Vela pulsar period was not what we expected. Further
observations over the next week or so confirmed this
(Fig.~\ref{fg:vela_glitch}) and led to the publication of the first
detection of a pulsar glitch \citep{rm69}.

As an aside, my wife Barbara also worked at Parkes in 1968 and 1969,
not on pulsars, but as a research assistant for John Bolton and Miller
Goss. As typing ``Manchester'' into the Astrophysics Data System for
1969 will reveal, she wrote two papers based on this work, one on
sources from the 1410 MHz Galactic plane survey \citep{man69} and one,
with Miller, on an 11cm map of the Vela region \citep{mg69}.

\section{Polarisation studies}
\begin{figure}
\resizebox{\hsize}{!}{\includegraphics{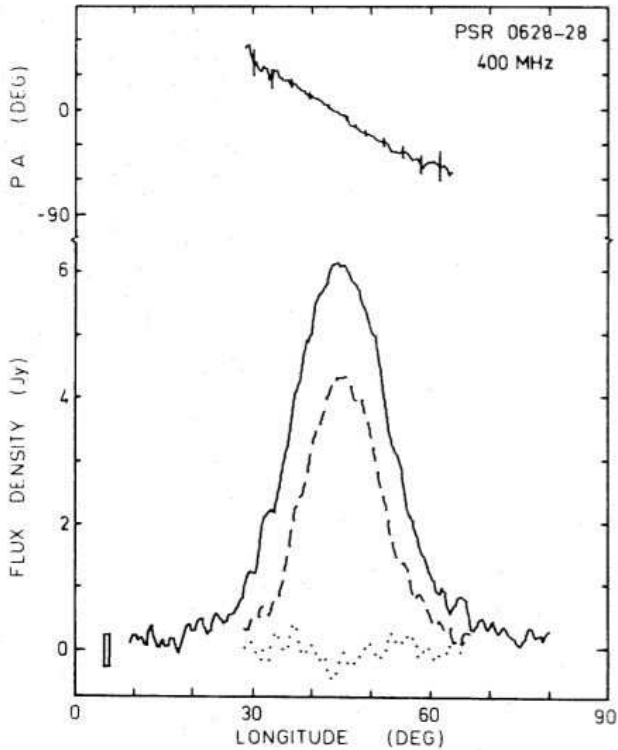}}
\caption{Mean pulse profile and polarisation properties of PSR
  B0628$-$28 at 400 MHz. The total-intensity pulse profile is shown
  as a full line, the linearly polarised component as a dashed
  line and the circularly polarised component as a dotted line. The
  postion angle of the linear polarisation is shown in the upper part
  of the figure. \citep{hmak77}.}\label{fg:B0628}
\end{figure}

As the Vela observations of \citet{rckm69} clearly demonstrated, a
large symmetric and steerable dish like Parkes is ideally suited to
polarisation studies -- its polarisaton reception properties are
essentially independent of pointing direction and are relatively easy to
calibrate. Great advantage was taken of this in subsequent
years. Systems providing simultaneous direct and cross products,
allowing a full description of the polarisation in a single
observation, were developed in the early 1970s by Peter McCulloch and
his colleagues \citep{mhak72}. These systems were used for an
extensive set of studies of the mean-pulse-profile polarisation of
pulsars at 400 MHz, 610 MHz and 1612 MHz \citep{hmak77,mhma78,mhm80}
as well as other observations of particular pulsars
\citep[e.g.,][]{mhak76}. Fig.~\ref{fg:B0628} illustrates the strong
linear polarisation and swing of position angle that is often observed. 

\begin{figure*}
\centerline{\includegraphics[width=140mm]{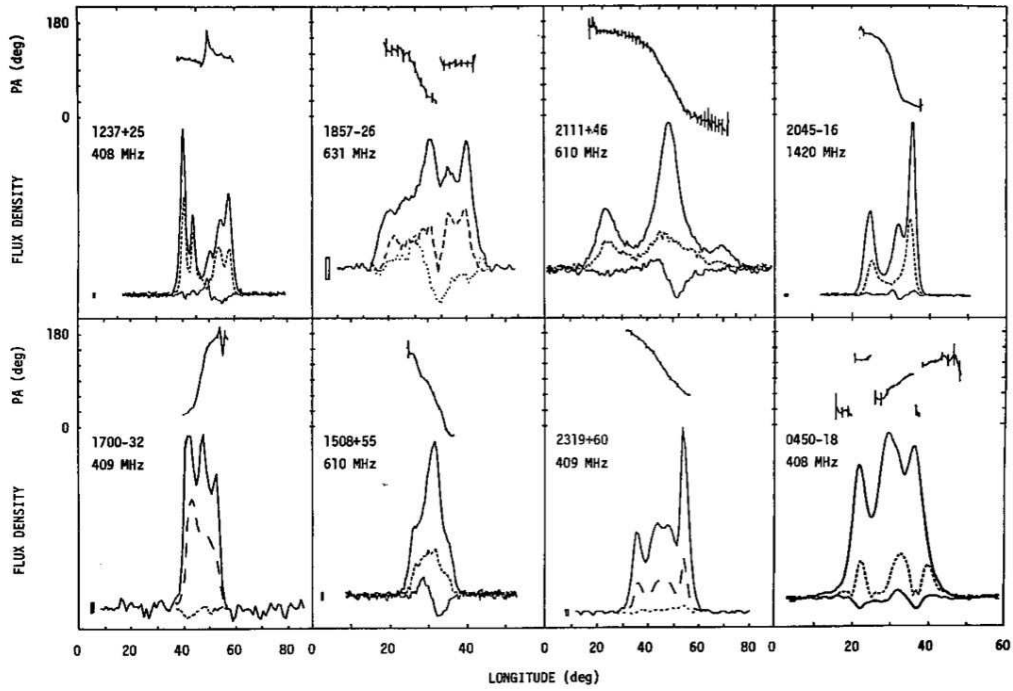}}
\caption{Mean pulse profiles for eight multi-component pulsars. The
  plot conventions are the same as in Fig.~\ref{fg:B0628}.
  \citep{lm88}.}\label{fg:poln}
\end{figure*}

Similar observations made at various observatories around the world
\citep[e.g.,][]{man71b,mhma78,mgs+81} were very important for our
understanding of radio pulsar beaming
\citep[e.g.,][]{ran83,lm88}. Fig.~\ref{fg:poln} shows polarisation
profiles for eight multicomponent pulsars illustrating the high degree
of linear polarisation often observed and the characteristic
``S''-shaped PA swings which closely conform to the rotating-vector
model. One case, PSR B1857$-$26, also illustrates the effect of
orthogonal-mode switching that is often observed in pulsar
polarisation data \citep[e.g.,][]{mth75,scr+84}. Although this
complicates the appearance of the PA curves, when it is allowed for,
the underlying PA swing is generally in accordance with the
rotating-vector model.

\begin{figure*}
\centerline{\includegraphics[height=140mm,angle=-90]{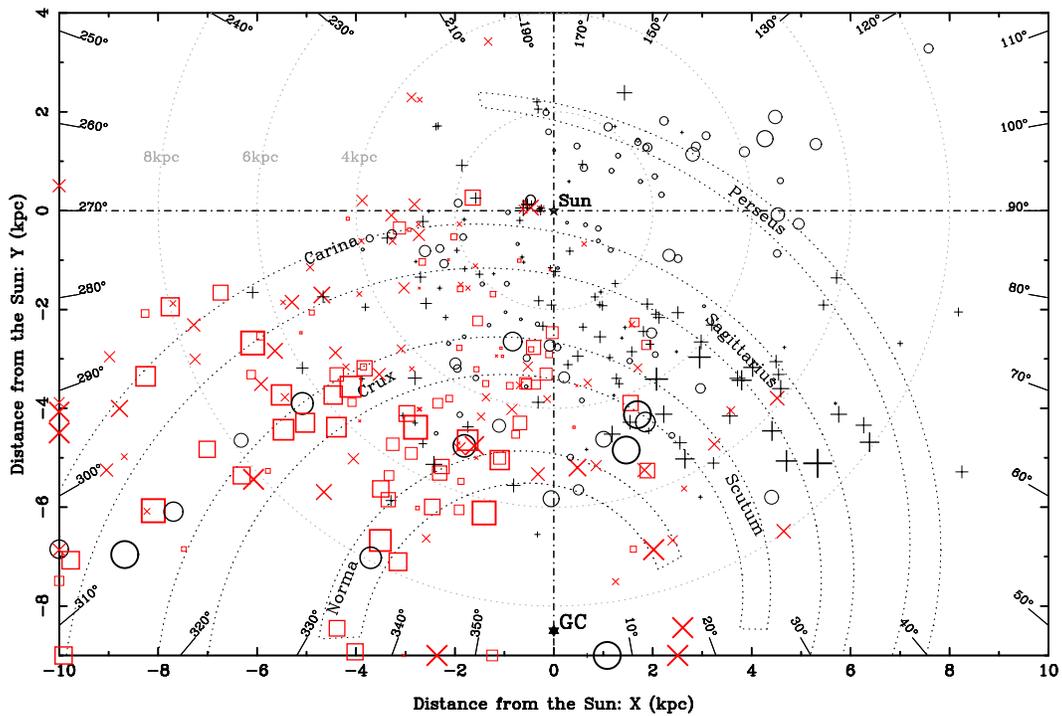}}
\caption{Pulsar rotation measures (RMs) projected on to the plane of the
  Galaxy. The size of the symbol is proportional to the magnitude of the
RM, + and $\times$ represent positive RMs, circles and squares represent
negative RMs and the red symbols ($\times$ and squares) represent RMs from
recent Parkes observations. \citep{hml+06} }\label{fg:gal_rm}
\end{figure*}

In later years at Parkes, many more studies of the mean pulse
polarisation of pulsars were undertaken
\citep[e.g.,][]{qmlg95,vdhm97,jkmg08,ymv+11} and these have contributed to a
growing understanding of the pulse emission mechanism. Pulsar
polarisation is also one of the best ways to investigate the structure
of the Galactic magnetic field. Pulsars' typically high linear
polarisation allows rotation measures (RMs) to be measured
relatively easily, they are distributed throughout the Galaxy at
approximately known distances, thereby allowing a tomographic analysis of the
field structure, and are point sources with no intrinsic Faraday
rotation. Fig.~\ref{fg:gal_rm} shows the distribution of RMs from
pulsar measurements, mostly done at Parkes, projected on to the plane of
the Galaxy. These results suggest that the Galactic disk
magnetic fields follow spiral arms with counter-clockwise
fields in the arms and clockwise fields in the inter-arm
regions \citep{hml+06}.

\section{Searches for pulsars at Parkes}
The Parkes radio telescope has an amazing record in searching for
pulsars, with the discovery of about two thirds of the known pulsars
to its credit. Put another way, this means that Parkes has discovered
twice as many pulsars as the rest of the world's telescopes put
together! There are several reasons for this great success. The first
is its southern location -- the Galactic Centre passes almost overhead
at Parkes and pulsars are more numerous in the inner parts of the
Galaxy. The second is the world-leading receiving systems that have
been and continue to be developed by our engineers and
astronomers. The outstanding example is the 20cm Multibeam receiver
and the associated filterbank back-end system, installed on Parkes in
1997 and responsible for about three-quarters of the Parkes
discoveries. Finally, the research groups involved in the more
successful pulsar searches at Parkes were very experienced in the
techniques of pulsar searching and discovery. 

The first successful pulsar search at Parkes was carried out by
\citet{kac+73}. They pioneered a two-dimensional Fourier analysis
technique that largely overcomes the effects of interstellar pulse
dispersion. This allowed the discovery of highly dispersed and distant
pulsars in the Galactic plane, most notably PSR B1641$-$45, one of the
most radio-luminous pulsars known with a dispersion measure (DM) of
478 cm$^{-3}$~pc and an estimated distance of 5.3 kpc.

\begin{figure*}[ht]
\centerline{\includegraphics[height=150mm,angle=-90]{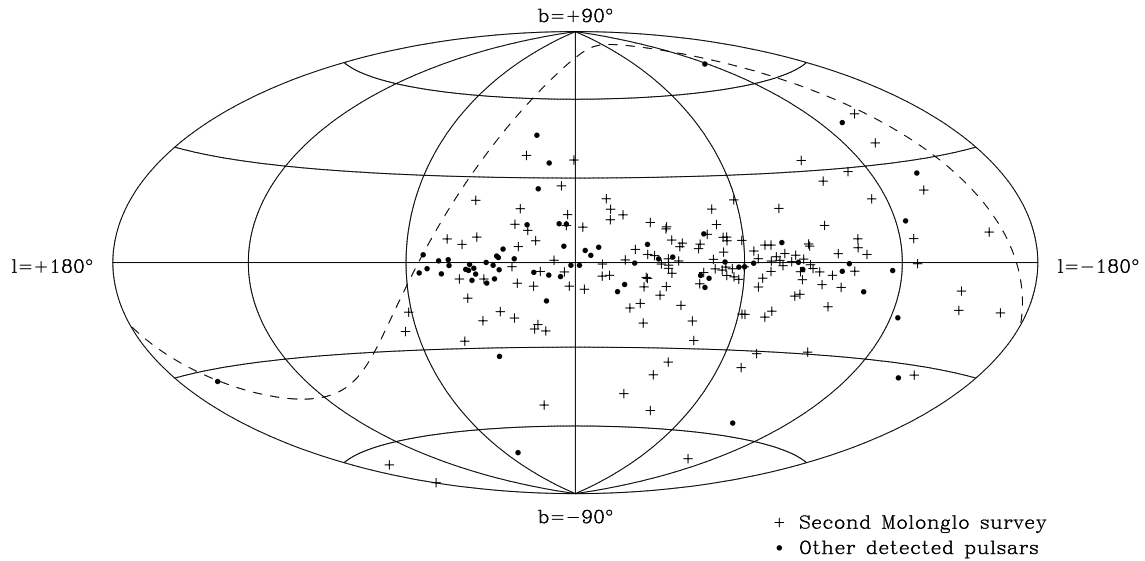}}
\caption{Distribution in Galactic coordinates of pulsars discovered by
  the Second Molonglo Pulsar Survey (+) and previously known pulsars
  detected by the survey (dot). The Galactic Centre is at $l=0$,
  $b=0$, and the dashed line corresponds to a declination of $+20\degr$,
  the northern limit of the survey.}\label{fg:mol2}
\end{figure*}

The next major survey to be undertaken was the Second Molonglo Pulsar
Survey, a joint effort between the University of Sydney and CSIRO
\citep{mlt+78}. With its large collecting area at 408 MHz and
$5\degr$-wide beam in declination, the Molonglo radio telescope was
used as the initial finding telescope, scanning the entire sky between
declinations of $-85\degr$ and $+20\degr$, approximately 67\% of the
celestial sphere. In order to maximise the sensitivity, we constructed
88 pre-amplifiers and installed them close to the output of each
feed bay, resulting in a system-equivalent flux density of 47 Jy, a
factor of 2.5 improvement over the original system. In addition, the
University of Sydney group developed a multibeam configuration that
gave an observation time of about 45\,s at the equator. Data were
processed using the CSIRO's Cyber 76 computer, located in Canberra,
and with the University of Sydney's Cyber 72 computer. Candidates from this
analysis were then observed at Parkes with a 408-MHz receiver by
scanning in declination along the Molonglo beam, and processing the
data in real time to determine an improved position, period and
dispersion measure for any confirmed pulsar. The survey was
outstandingly successful, discovering 155 previously unknown pulsars,
more than doubling the number of pulsars known at the time. The
distribution on the celestial sphere of the 155 newly discovered and
the 69 previously known pulsars detected in the survey is shown in
Fig.~\ref{fg:mol2}. Most of the new discoveries are at relatively low
Galactic latitudes and in the fourth Galactic quadrant, that is,
concentrated in the disk of the Galaxy and towards the Galactic
Centre.

\begin{figure}
\resizebox{\hsize}{!}{\includegraphics{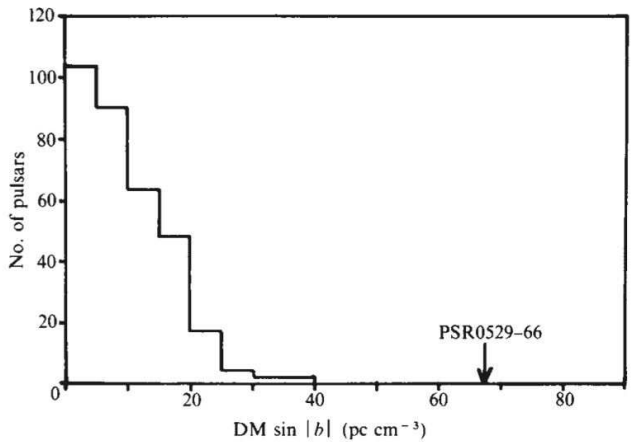}}
\caption{Distribution of DM\,$\sin |b|$, where $b$ is Galactic
  latitude, for the 330 pulsars known at the time of the discovery of
  PSR B0529-66. This pulsar is well outside the distribution of this
  parameter (effectively the ``z-component'' of DM) for Galactic
  pulsars, indicating its extra-Galactic
  location. \citep{mhah83}}\label{fg:MagCld}
\end{figure}

Another notable Parkes achievement was the discovery by \citet{mhah83}
of the first extra-galactic pulsar. Using a dual-channel receiver at
645 MHz and a multi-channel dedispersion system, they searched 29 beam
areas toward the Large Magellanic Cloud (LMC), spending two hours on
each position. From a Fourier analysis of the data and subsequent
folding at candidate periods and DMs, they confirmed one pulsar, PSR
B0529-66, with a period of 0.975\,s and a DM of 125
cm$^{-3}$\,pc. Fig.~\ref{fg:MagCld} shows that this pulsar lies well
outside the distribution of DM\,$\sin |b|$ for Galactic pulsars which,
with its position, confirmed its location in the LMC. This search was
extended with an improved receiver system by \citet{mmh+91} who
discovered four pulsars, one in the Small Magellanic Cloud (SMC), two
in the LMC and one foreground pulsar. Further timing at Parkes by
\citet{kjb+94} showed that the SMC pulsar, now known as PSR
J0045$-$7319, is in an eccentric 51-day binary orbit around a massive
star, which was optically identified as a B-star by
\citet{bbs+95}. After the advent of the 20cm Multibeam receiver,
\citet{ckm+01} made a much deeper survey of the SMC, discovering one
pulsar likely to be in the SMC and also found a pulsar in the LMC in a
separate search. A much more complete survey of both Magellanic Clouds
was undertaken by \citet{mfl+06}; they discovered a further nine
pulsars associated with the LMC and three with the SMC. Together with
two pulsars discovered at X-ray wavelengths \citep{shh84,mgz+98} this
brought the total number of pulsars known in the Magellanic Clouds to
20. These remain the only extra-Galactic pulsars known.

\begin{figure*}[ht]
\centerline{\includegraphics[width=120mm]{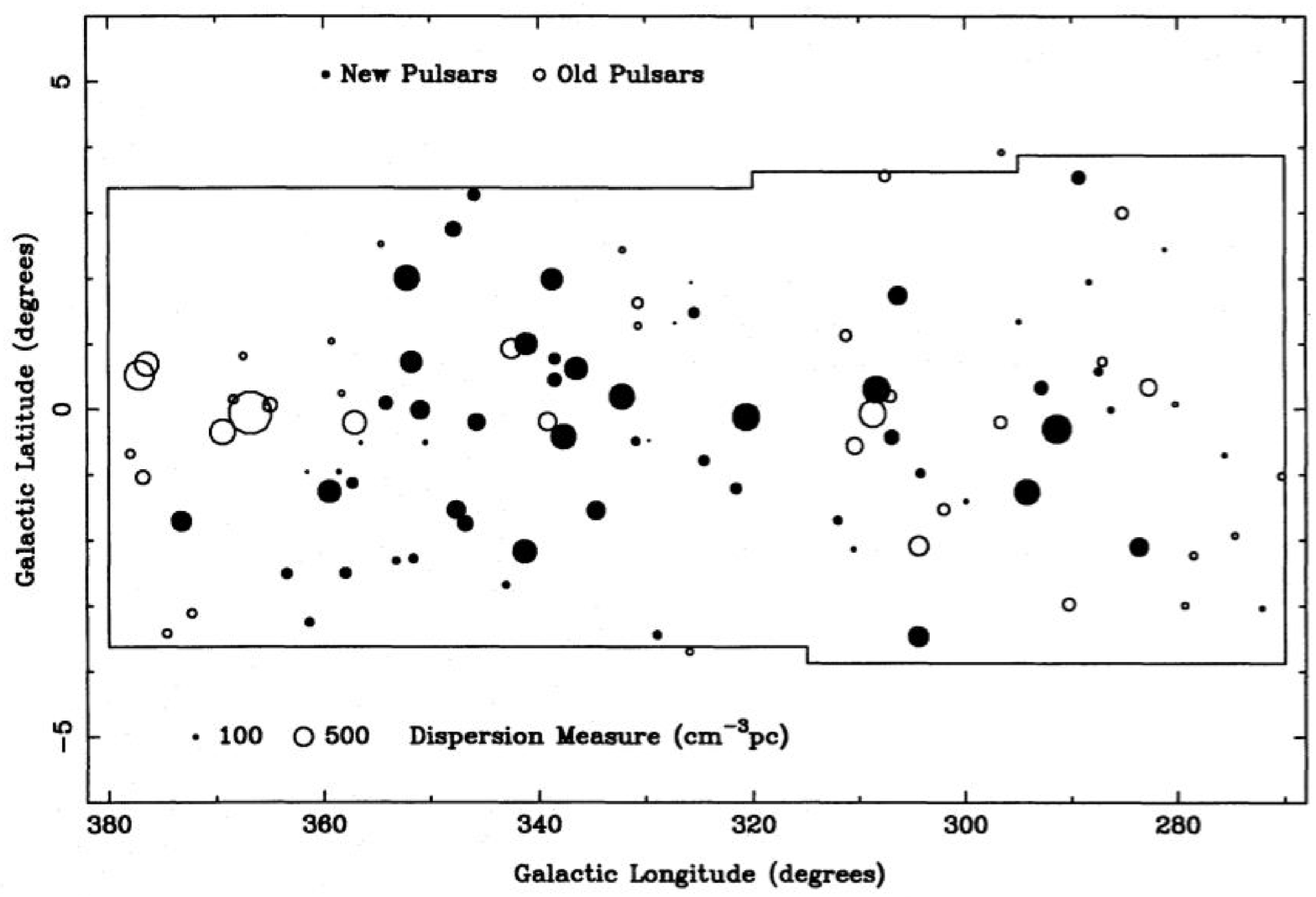}}
\caption{Distribution in Galactic coordinates of the 100 pulsars
  detected in the Parkes Southern Galactic Plane Survey. New
  discoveries are marked with filled circles and the areas of the
  circles are proportional to the pulsar dispersion
  measure. \citep{jlm+92}}\label{fg:pks1}
\end{figure*}

After the Second Molonglo Pulsar Survey, the next major pulsar survey
at Parkes was the Parkes Southern Galactic Plane Survey
\citep{jlm+92}. This survey was restricted to $\pm3\degr$ of the
Galactic Plane and was the first of the major 1400 MHz (20cm) surveys,
complementing a similar survey of the northern Galactic plane at
Jodrell Bank \citep{clj+92}. The higher frequency allowed deeper
penetration into the Galactic disk before the effects of interstellar
dispersion and scattering became a limiting factor. The survey was
very successful with 46 previously unknown pulsars discovered and a
total of 100 pulsars detected. Half of the new discoveries
have DMs greater than 300 cm$^{-3}$\,pc, illustrating the much greater
sensitivity of this survey to high-DM pulsars. 
(Fig.~\ref{fg:pks1}) 

The most interesting pulsar discovered in this survey is undoubtedly
PSR B1259-63 \citep{jml+92}. This pulsar had the shortest period of all
those discovered in the survey, 47~ms, but more than that, it turned
out to be in a long-period and highly eccentric orbit around a massive
star, later optically identified as SS~2883, a 10th magnitude Be
star. Initially, the orbital period and eccentricity were not well
defined, but later measurements \citep{jml+94} showed that they were
3.4 years and 0.87, respectively. The pulsar is eclipsed by the
emission-line disk surrounding the Be star every periastron and the
system also emits transient radio \citep{jmmc99}, X-ray
\citep{ktn+95,gtp+95}, gamma-ray \citep{aaa+11} and TeV
\citep{aaa+09l} emission as the pulsar ploughs through the disk around
periastron.

The discovery by \citet{lbm+87} of a millisecond pulsar (MSP) in the
globular cluster M~28 and subsequent discoveries in other clusters
showed that the dense stellar environment of the cores of such
clusters are conducive to the formation of MSPs. The first Parkes
discovery of a pulsar associated with a globular cluster was by
\citet{mld+90} who found a 5.75-ms pulsar in 47 Tucanae. Subsequent
observations of 47 Tucanae \citep{mlr+91,rlm+95,clf+00} detected a
total of 20 MSPs in this one cluster, giving it the highest
concentration of MSPs known -- until the detection of 21 MSPs in
Terzan~5 by \citet{rhs+05} which brought the number of pulsars
known in that cluster to 24. Timing observations of the 47 Tucanae
pulsars have revealed many interesting results, for example, the first
detection of ionised intra-cluster gas in a globular cluster
\citep{fkl+01} and the proper motion of the cluster
\citep{fcl+01}. Parkes searches of other clusters have resulted in a
dozen or so MSP discoveries \citep[e.g.,][]{pdm+03}.

\begin{figure*}[ht]
\centerline{\includegraphics[width=160mm]{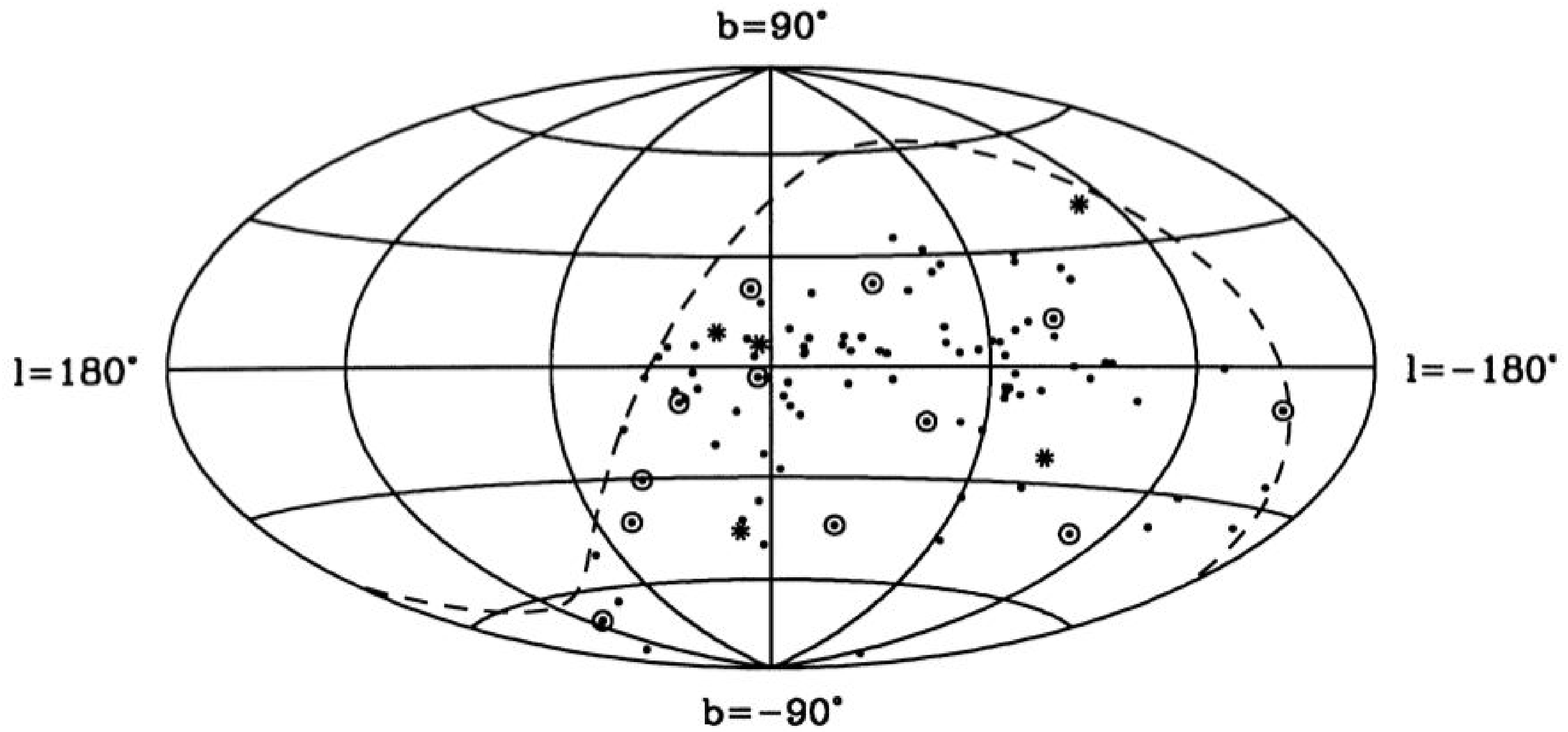}}
\caption{Distribution in Galactic coordinates of the 101 pulsars
  discovered in the Parkes Southern Pulsar Survey. Solitary MSPs are
  indicated by an asterisk, binary MSPs by a point with a circle and
  other pulsars by a point. 
  \citep{lml+98}}\label{fg:pks70}
\end{figure*}

Although searching globular clusters is certainly a productive way to
find MSPs, for many applications it was (and is) important to find
MSPs in the Galactic field. The gravitational interactions with other
cluster stars introduce perturbations into the observed pulse periods,
and, while these are intrinisically interesting, they limit the
application of these pulsars to many precision timing projects. Also,
since MSPs have a very long lifetime compared to ``normal'' pulsars,
they can be expected to lie at large distances from the Galactic plane
and hence at relatively high Galactic latitudes. With these ideas in
mind, a survey of the entire Southern sky with rapid sampling to give
sensitivity to MSPs was planned. To cover such a
large area, it was necessary to use a relative low frequency and a
receiver system centred at 436 MHz was constructed \citep{mld+96}. The
survey was very successful with 101 new discoveries including 17 MSPs
and a total of 298 pulsars detected \citep{lml+98}. As
Fig.~\ref{fg:pks70} shows, most of the MSPs were at high Galactic
latitudes, in accordance with expectations. 

\begin{figure*}[ht]
\centerline{\includegraphics[height=100mm,angle=-90]{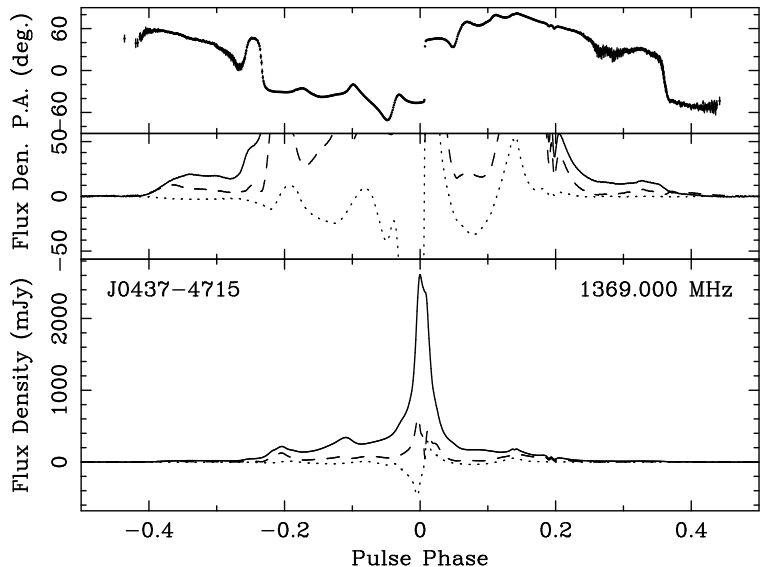}}
\caption{Polarisation profiles for the 5.75-ms pulsar J0437$-$4715 at
  at (or around) 1400 MHz. The lower part gives the pulse profile for
  total intensity (solid line), linearly polarized intensity
  (dashed line) and circularly polarized intensity (dotted
  line). In the middle part the vertical scale is expanded by a factor
  of 20 to show the low-level details of the profiles, and the upper
  part gives the position angle of the linearly polarized
  emission. \citep{ymv+11}}\label{fg:J0437poln}
\end{figure*}

The most important discovery from this survey was the very bright and
nearby MSP PSR J0437$-$4715 \citep{jlh+93}. This pulsar has a period
of 5.75~ms, DM of just 2.64 cm$^{-3}$\,pc and a mean flux density at
1400 MHz of about 140 mJy, an order of magnitude greater than any
other MSP. Precise timing observations of this pulsar have allowed
measurement of the ``annual orbital parallax'', the apparent change in
the projected orbit size due to the annual motion of the Earth, giving
a constraint on the longitude of the ascending node and hence allowing
an independent test of the predictions of general relativity
\citep{vbb+01}. These measurements also gave a precise measurement of
the pulsar distance, $157.0\pm 2.4$~pc, based on the apparent change in
orbital period due to transverse motion of the system
\citep{vbv+08}. This timing distance is fully consistent with the even
more precise VLBI measurement of \citet{dvtb08}, $156.3\pm1.3$~pc. The
extraordinary signal/noise ratio obtainable in observations of this
pulsar can be seen in Fig.~\ref{fg:J0437poln} which shows the mean
pulse profile polarisation at 1400 MHz \citep{ymv+11}.

\begin{figure*}[ht]
\centerline{\includegraphics[height=160mm,angle=-90]{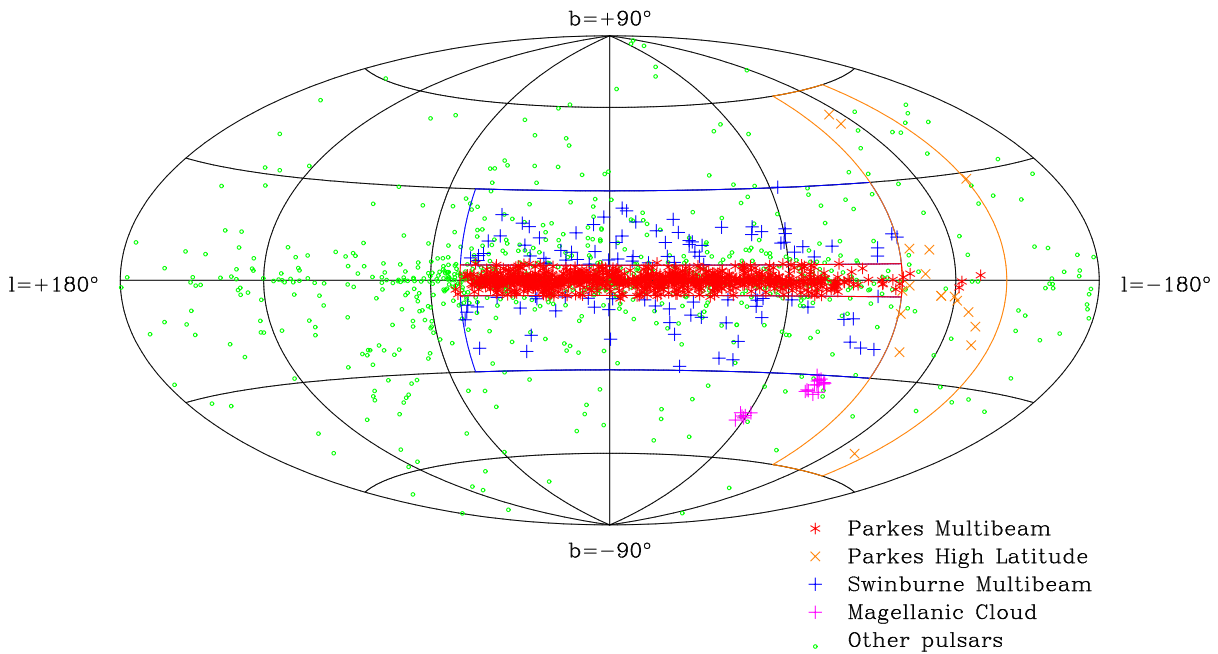}}
\caption{Galactic distribution of all known pulsars except those
  associated with globular clusters. Pulsars discovered in the various
  surveys with the Parkes 20cm Multibeam system (prior to 2010) are
  marked. (Data from the ATNF Pulsar Catalogue, V1.43.)}\label{fg:pksmb}
\end{figure*}

The 1996 commissioning of the 20cm Parkes Multibeam receiver \citep{swb+96}
resulted in a quantum leap in the efficiency and effectiveness of
pulsar surveys at Parkes. Not only did the 13 beams allow searches to
cover the sky more than an order of magnitude faster, but the low
system temperature of the receivers, about 21~K, gave excellent
sensitivity. A large number of pulsar surveys have been undertaken
with this receiver. Fig.~\ref{fg:pksmb} shows the Galactic
distribution of all currently known pulsars with those found in
searches using the Multibeam receiver highlighted. By far the most
successful survey (so far) is the Parkes Multibeam Pulsar Survey which
covered a $\pm5\degr$ strip along the Galactic plane with
unprecedented sensitivity. The six main papers from the survey
\citep{mlc+01,mhl+02,kbm+03,hfs+04,fsk+04,lfl+06} reported the
discovery of 759 pulsars, but various reprocessing efforts
\citep[eg.,][]{kel+09} have now raised that to 785 discoveries. In
total, this single survey has detected 1065 pulsars, more than half
the total number known, giving a wonderful database for studies of the
Galactic distribution and evolution of pulsars
\citep[e.g.,][]{lfl+06,fk06}.

\begin{figure}
\resizebox{\hsize}{!}{\includegraphics{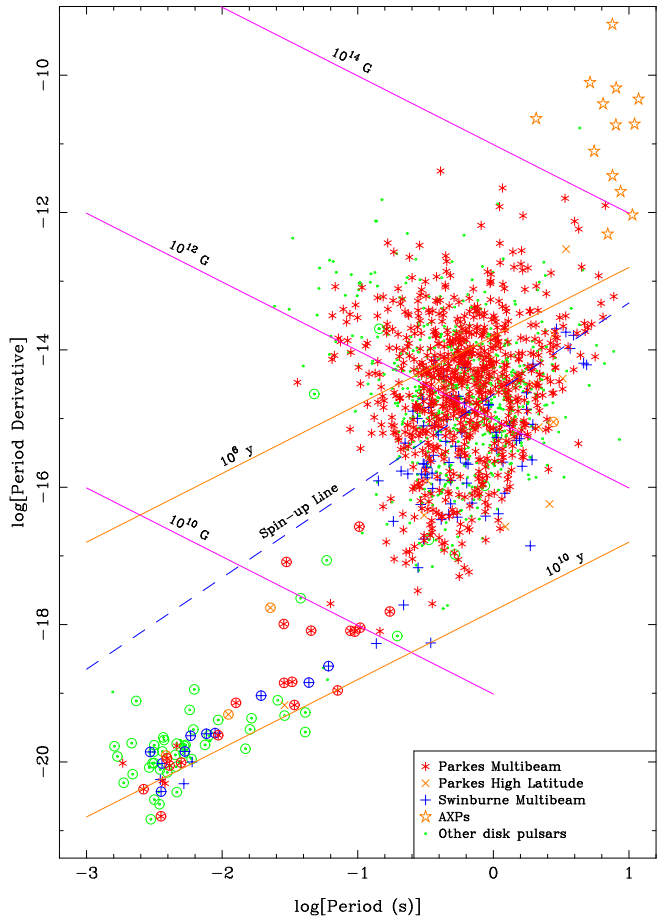}}
\caption{Period -- period derivative diagram for pulsars discovered in
  the three main Parkes 20cm multibeam surveys, anomalous X-ray
  pulsars and other pulsars in the Galactic disk. Binary pulsars are
  indicated by a circle around the symbol. Lines of constant
  surface-dipole magnetic field and constant characteristic age are
  shown. The spin-up (dashed) line indicates the minimum period that a
  pulsar can reach, given certain simplifying assumptions
  \citep[see][]{bv91}, by accretion of matter from a companion
  star.}\label{fg:ppdot}
\end{figure}

The plot of the rate of period increase $\dot P$ versus period $P$ is
one of the basic diagrams of pulsar astrophysics. It reveals the
various families of pulsars, separating, for example, the millisecond
pulsars and the anomalous X-ray pulsars (AXPs)\footnote{As in the ATNF
  Pulsar Catalogue
  \citep[][www.atnf.csiro.au/research/pulsar/psrcat]{mhth05},
  the term ``AXP'' is taken to include soft gamma-ray repeaters
  (SGRs), and to represent the group of pulsars often labelled
  ``magnetars''} from the ``normal'' pulsars. Fig.~\ref{fg:ppdot}
shows this diagram, with the pulsars discovered in the main Parkes
multibeam surveys marked. Millisecond pulsars are concentrated in the
lower-left corner of the diagram with very long characteristic ages,
often greater than $10^9$~years, and relatively weak implied dipole
magnetic fields. In contrast, AXPs are located in the upper-right part
of the diagram with ages typically around $10^4$~years and very strong
implied magnetic fields, $\sim 10^{15}$~G ($10^{11}$~T), hence the
term ``magnetar'' for these pulsars. The vast bulk of normal pulsars
have characteristic ages of between $10^6$ and $10^7$~years and
surface dipole fields of $\sim 10^{12}$~G. One of the notable
achievements of the Parkes Multibeam Pulsar Survey was the discovery of
a whole new class of pulsars with implied magnetic field strengths
and, in some cases pulse periods, approaching those typical of the
magnetars \citep[e.g.,][]{msk+03}. In Fig.~\ref{fg:ppdot} these are
located just below the $10^{14}$~G line. These pulsars raise
intriguing issues about why magnetars differ from radio pulsars and
whether or not there is an evolutionary relationship between the two
groups \citep[e.g.,][]{elk+11}.

\section{The Double Pulsar}

Without doubt, the ``Jewel in the Crown'' of the Parkes multibeam
surveys was the discovery of the Double Pulsar, PSR J0737$-$3039A/B
\citep{bdp+03,lbk+04} in the Parkes High-Latitude Survey. PSR
J0737$-$3039A, seen as a circled orange $\times$ just below the
$10^{10}$~G line in Fig.~\ref{fg:ppdot}, is part of a large group of
mildly recycled pulsars discovered in the multibeam surveys. Most of the
binary systems in this group are thought to be double-neutron-star
systems, where the companion star, as well as the pulsar, is a neutron
star. However, PSR~J0737$-$3039A/B was distinguished both by its very
short orbital period, 2.4~hours, and by the fact that it was the only
known double-neutron-star system where pulses are observed from both
neutron stars, thus making it the first, and to date only,
double-pulsar system known. Pulsar A is a mildly recycled pulsar with
a pulse period of 22.7~ms, whereas Pulsar B has a long period, 2.77~s
and a much stronger magnetic field -- it can be seen in
Fig.~\ref{fg:ppdot} just above the $10^{12}$~G line near the
right-hand edge of the pulsar distribution. Pulsar B was the
second-born of the two pulsars and, during its pre-supernova
evolution, spun up or recycled Pulsar A to very close to its current
short period. Another distinguishing feature of the Double Pulsar
system is its orbital inclination. By great good fortune -- or some
aspect of pulsar astrophysics that we don't currently understand --
the orbit is almost exactly edge-on to us, allowing investigation of a
number of interesting and sometimes unique phenomena. 

The Double Pulsar system, named by Science magazine as one of the top ten
scientific break-throughs of 2004, has proved to be a remarkable
laboratory for the investigation of relativistic astrophysics. Because
of the very short orbital period, relativistic perturbations to the
orbit dynamics are easily observed, with five measured so far
\citep{ksm+06}. These, together with the mass ratio for the two stars,
determined from their observed Doppler shifts and unique to the Double
Pulsar, give four independent checks on the gravitational theory used
to interpret them, namely Einstein's general theory of relativity
(GR). Already in 2006, these showed that GR is accurate at the 0.05\%
level in the strong-field regime, easily surpassing tests based on the
original Hulse-Taylor binary pulsar \citep{wt05}.

The Double Pulsar has other claims to fame. Some of the most important
arise from the fact that the orbit is seen nearly edge-on; the orbital
inclination measured from pulse timing is $\sim 88.7\degr$
\citep{ksm+06}. This leads to eclipses of the radiation from Pulsar A
by the magnetosphere of Pulsar B, giving a unique probe into pulsar
magneto-hydrodynamics. The eclipses last only 30 seconds in the
2.4-hour orbit, showing that most of B's magnetosphere is blown away
by the relativistic wind from Pulsar A. More than that, they are
modulated at half the period of pulsar B \citep{mll+04}. This
modulation has been fitted by a model in which the eclipses are due to
synchrotron absorption occurring in the closed-field-line torus of
Pulsar B \citep{lt05}. \citet{bkk+08} have shown that the exact form
of the eclipse curve changes in time because of the relativistic
precession of B's spin axis, giving yet another independent test of
GR.

\citet{mkl+04} detected another fascinating aspect of the interaction
between Pulsar A and Pulsar B -- the emission from Pulsar B is
modulated at the rotation frequency of Pulsar A (Doppler-shifted by
the relative motion of A and B around their orbit). This implies that
some aspect of the magneto-hydrodynamic wind from Pulsar A is
affecting the magnetosphere of Pulsar B to such an extent that it shuts off,
or maybe redirects, B's pulsed emission every A period. 

The precession of B's spin axis has had another interesting
consequence -- since observations of the system began in 2003, the
pulsed emission from Pulsar B has been getting steadily weaker and its
pulse profile has been getting broader \citep{bpm+05}. In 2008 Pulsar B became
undetectable so the system is no longer a double pulsar! However, the
prediction based on a model for the emission beam is that Pulsar B will
reappear before 2035 and maybe as early as 2014 \citep{pmk+10}.

More details on the Double Pulsar system are in the paper
by Marta Burgay in these Proceedings. 

\section{Pulsar Timing}
The importance of pulsars rests largely on their incredibly stable
pulse periods. This stability is a direct result of the large mass,
$\sim 1.4$~M$_\odot$, and tiny radius, $\sim 15$~km, of neutron
stars. These properties make it extremely difficult to change the spin
rate of a neutron star or, equivalently, the pulse period. Pulsar
timing has a huge range of applications, ranging from investigations
of the interior of neutron stars to the detection of gravitational
waves. To determine accurate pulse periods -- and they way in which
they vary -- it is necessary to make frequent observations over long
time intervals, typically several years or even decades. Each
observation yields a pulse time of arrival (ToA) at the telescope. A
series of these ToAs is compared with the predictions of a model for
the pulsar, including its position, proper motion and sometimes annual
parallax, the pulse frequency and its derivatives and, if necessary,
the parameters of its binary motion. The differences between the
observed ToAs and the model predictions are known as timing
residuals. Systematic deviations of these timing residuals from zero
indicate either errors in the assumed pulsar model or the existence of
unmodelled phenomena affecting the observed pulse period. Errors in
the timing model can be determined by a least-squares fit to the
residual variations, thereby improving the model. If unmodelled
residual variations can be identified, then terms for these can (in
principle) be included in the model.

The Parkes telescope has provided the platform for many pulsar timing
investigations dating right back to 1968 -- some of these have already
been mentioned above. Many were follow-up timing observations of
pulsars discovered in the various pulsar searches at Parkes. Such
observations are vital to understanding the properties of newly
discovered pulsars and have of course led to many important results -
the timing of the Double Pulsar is just one example. Over the past few
years, our main project has been the Parkes Pulsar Timing Array
(PPTA). This project differs from most other pulsar timing projects in
that it involves observing a relatively large sample of MSPs at
regular intervals over a long data span. The aim is to detect tiny
signals that affect all pulsars in the array in a correlated way. Our
main objectives are the direct detection of gravitational waves and
establishment of a pulsar-based timescale, but there are many secondary
objectives. For the PPTA we observe 20 MSPs at three frequency bands,
10cm, 20cm and 50cm, and at intevals of 2 -- 3 weeks. Regular
observations commenced in 2004 and they are still continuing. The PPTA
project is described in more detail in the paper by George Hobbs in
these Proceedings.

\section{Conclusions}
The Parkes radio telescope has been a wonderful instrument for the
discovery and study of pulsars and I feel priviliged to have been part
of that history. Many important contributions have been made and some
Parkes pulsar papers are among the most highly cited Australian
astronomy publications. Some of this success results from the location
of Parkes in the Southern Hemisphere with its favourable view of our
galaxy, the Milky Way. Of course, it also rests on the foresight and
determination of Taffy Bowen and others to realise the dream of a
giant fully steerable ``dish'' in Australia. But, after that, most of
the credit belongs to the great collaborations between our
engineers and scientists to conceive, design, build and operate
world-leading receiving and signal-processing systems. It is a tribute
to them that Parkes has been and remains at the forefront of
international radio astronomy research, especially in the field of
pulsar astronomy. With continued development of instrumentation and
continued support from CASS, we can look forward to many more decades
of great science!

\section*{Acknowledgments}
I thank my colleagues for their invaluable contributions to the
various collaborative projects described in this review. The Parkes
radio telescope is part of the Australia Telescope National Facility
which is funded by the Commonwealth of Australia for operation as a
National Facility managed by CSIRO.
 
%\bibliographystyle{apj}
%\bibliography{journals,modrefs,psrrefs,crossrefs}

\begin{thebibliography}{81}
\expandafter\ifx\csname natexlab\endcsname\relax\def\natexlab#1{#1}\fi

\bibitem[{{Abdo} {et~al.}(2011){Abdo}, {Ackermann}, {Ajello}, {Allafort},
  {Ballet}, {Barbiellini}, {Bastieri}, {Bechtol}, \& et~al.}]{aaa+11}
{Abdo}, A.~A., {Ackermann}, M., {Ajello}, M., {Allafort}, A., {Ballet}, J.,
  {Barbiellini}, G., {Bastieri}, D., {Bechtol}, K., \& et~al. 2011, ApJ, 736,
  L11

\bibitem[{{Aharonian} {et~al.}(2009){Aharonian}, {Akhperjanian}, {Anton},
  {Barres de Almeida}, {Bazer-Bachi}, {Becherini}, {Behera}, {Bernl{\"o}hr}, \&
  et~al.}]{aaa+09l}
{Aharonian}, F., {Akhperjanian}, A.~G., {Anton}, G., {Barres de Almeida}, U.,
  {Bazer-Bachi}, A.~R., {Becherini}, Y., {Behera}, B., {Bernl{\"o}hr}, K., \&
  et~al. 2009, A\&A, 507, 389

\bibitem[{Bell {et~al.}(1995)Bell, Bessell, Stappers, Bailes, \&
  Kaspi}]{bbs+95}
Bell, J.~F., Bessell, M.~S., Stappers, B.~W., Bailes, M., \& Kaspi, V.~M. 1995,
  ApJ, 447, L117

\bibitem[{Bhattacharya \& {van den Heuvel}(1991)}]{bv91}
Bhattacharya, D. \& {van den Heuvel}, E. P.~J. 1991, Phys. Rep., 203, 1

\bibitem[{{Breton} {et~al.}(2008){Breton}, {Kaspi}, {Kramer}, {McLaughlin},
  {Lyutikov}, {Ransom}, {Stairs}, {Ferdman}, {Camilo}, \& {Possenti}}]{bkk+08}
{Breton}, R.~P., {Kaspi}, V.~M., {Kramer}, M., {McLaughlin}, M.~A., {Lyutikov},
  M., {Ransom}, S.~M., {Stairs}, I.~H., {Ferdman}, R.~D., {Camilo}, F., \&
  {Possenti}, A. 2008, Science, 321, 104

\bibitem[{{Burgay} {et~al.}(2003){Burgay}, {D'Amico}, {Possenti}, {Manchester},
  {Lyne}, {Joshi}, {McLaughlin}, {Kramer}, {Sarkissian}, {Camilo}, {Kalogera},
  {Kim}, \& {Lorimer}}]{bdp+03}
{Burgay}, M., {D'Amico}, N., {Possenti}, A., {Manchester}, R.~N., {Lyne},
  A.~G., {Joshi}, B.~C., {McLaughlin}, M.~A., {Kramer}, M., {Sarkissian},
  J.~M., {Camilo}, F., {Kalogera}, V., {Kim}, C., \& {Lorimer}, D.~R. 2003,
  Nature, 426, 531

\bibitem[{{Burgay} {et~al.}(2005){Burgay}, {Possenti}, {Manchester}, {Kramer},
  {McLaughlin}, {Lorimer}, {Stairs}, {Joshi}, {Lyne}, {Camilo}, {D'Amico},
  {Freire}, {Sarkissian}, {Hotan}, \& {Hobbs}}]{bpm+05}
{Burgay}, M., {Possenti}, A., {Manchester}, R.~N., {Kramer}, M., {McLaughlin},
  M.~A., {Lorimer}, D.~R., {Stairs}, I.~H., {Joshi}, B.~C., {Lyne}, A.~G.,
  {Camilo}, F., {D'Amico}, N., {Freire}, P.~C.~C., {Sarkissian}, J.~M.,
  {Hotan}, A.~W., \& {Hobbs}, G.~B. 2005, ApJ, 624, L113

\bibitem[{{Camilo} {et~al.}(2000){Camilo}, {Lorimer}, {Freire}, {Lyne}, \&
  {Manchester}}]{clf+00}
{Camilo}, F., {Lorimer}, D.~R., {Freire}, P., {Lyne}, A.~G., \& {Manchester},
  R.~N. 2000, ApJ, 535, 975

\bibitem[{Clifton {et~al.}(1992)Clifton, Lyne, Jones, McKenna, \&
  Ashworth}]{clj+92}
Clifton, T.~R., Lyne, A.~G., Jones, A.~W., McKenna, J., \& Ashworth, M. 1992,
  MNRAS, 254, 177

\bibitem[{{Crawford} {et~al.}(2001){Crawford}, {Kaspi}, {Manchester}, {Lyne},
  {Camilo}, \& {D'Amico}}]{ckm+01}
{Crawford}, F., {Kaspi}, V.~M., {Manchester}, R.~N., {Lyne}, A.~G., {Camilo},
  F., \& {D'Amico}, N. 2001, ApJ, 553, 367

\bibitem[{{Deller} {et~al.}(2008){Deller}, {Verbiest}, {Tingay}, \&
  {Bailes}}]{dvtb08}
{Deller}, A.~T., {Verbiest}, J.~P.~W., {Tingay}, S.~J., \& {Bailes}, M. 2008,
  ApJ, 685, L67

\bibitem[{{Espinoza} {et~al.}(2011){Espinoza}, {Lyne}, {Kramer}, {Manchester},
  \& {Kaspi}}]{elk+11}
{Espinoza}, C.~M., {Lyne}, A.~G., {Kramer}, M., {Manchester}, R.~N., \&
  {Kaspi}, V.~M. 2011, ApJ, 741, L13

\bibitem[{{Faucher-Gigu{\`e}re} \& {Kaspi}(2006)}]{fk06}
{Faucher-Gigu{\`e}re}, C.-A. \& {Kaspi}, V.~M. 2006, ApJ, 643, 332

\bibitem[{{Faulkner} {et~al.}(2004){Faulkner}, {Stairs}, {Kramer}, {Lyne},
  {Hobbs}, {Possenti}, {Lorimer}, {Manchester}, {McLaughlin}, {D'Amico},
  {Camilo}, \& {Burgay}}]{fsk+04}
{Faulkner}, A.~J., {Stairs}, I.~H., {Kramer}, M., {Lyne}, A.~G., {Hobbs}, G.,
  {Possenti}, A., {Lorimer}, D.~R., {Manchester}, R.~N., {McLaughlin}, M.~A.,
  {D'Amico}, N., {Camilo}, F., \& {Burgay}, M. 2004, MNRAS, 355, 147

\bibitem[{Freire {et~al.}(2001a)Freire, Kramer, Lyne, Camilo, Manchester, \&
  D'Amico}]{fkl+01}
Freire, P.~C., Kramer, M., Lyne, A.~G., Camilo, F., Manchester, R.~N., \&
  D'Amico, N. 2001a, ApJ, 557, L105

\bibitem[{{Freire} {et~al.}(2001b){Freire}, {Camilo}, {Lorimer}, {Lyne},
  {Manchester}, \& {D'Amico}}]{fcl+01}
{Freire}, P.~C., {Camilo}, F., {Lorimer}, D.~R., {Lyne}, A.~G., {Manchester},
  R.~N., \& {D'Amico}, N. 2001b, MNRAS, 326, 901

\bibitem[{Grove {et~al.}(1995)Grove, Tavani, Purcell, Kurfess, Strickman, \&
  Arons}]{gtp+95}
Grove, E., Tavani, M., Purcell, W.~R., Kurfess, J.~D., Strickman, M.~S., \&
  Arons, J. 1995, ApJ, 447, L113

\bibitem[{Hamilton {et~al.}(1977)Hamilton, McCulloch, Ables, \&
  Komesaroff}]{hmak77}
Hamilton, P.~A., McCulloch, P.~M., Ables, J.~G., \& Komesaroff, M.~M. 1977,
  MNRAS, 180, 1

\bibitem[{Han {et~al.}(2006)Han, Manchester, Lyne, Qiao, \& van
  Straten}]{hml+06}
Han, J.~L., Manchester, R.~N., Lyne, A.~G., Qiao, G.~J., \& van Straten, W.
  2006, ApJ, 642, 868

\bibitem[{Hewish {et~al.}(1968)Hewish, Bell, Pilkington, Scott, \&
  Collins}]{hbp+68}
Hewish, A., Bell, S.~J., Pilkington, J. D.~H., Scott, P.~F., \& Collins, R.~A.
  1968, Nature, 217, 709

\bibitem[{{Hobbs} {et~al.}(2004){Hobbs}, {Faulkner}, {Stairs}, {Camilo},
  {Manchester}, {Lyne}, {Kramer}, {D'Amico}, {Kaspi}, {Possenti}, {McLaughlin},
  {Lorimer}, {Burgay}, {Joshi}, \& {Crawford}}]{hfs+04}
{Hobbs}, G., {Faulkner}, A., {Stairs}, I.~H., {Camilo}, F., {Manchester},
  R.~N., {Lyne}, A.~G., {Kramer}, M., {D'Amico}, N., {Kaspi}, V.~M.,
  {Possenti}, A., {McLaughlin}, M.~A., {Lorimer}, D.~R., {Burgay}, M., {Joshi},
  B.~C., \& {Crawford}, F. 2004, MNRAS, 352, 1439

\bibitem[{{Johnston} {et~al.}(2008){Johnston}, {Karastergiou}, {Mitra}, \&
  {Gupta}}]{jkmg08}
{Johnston}, S., {Karastergiou}, A., {Mitra}, D., \& {Gupta}, Y. 2008, MNRAS,
  388, 261

\bibitem[{Johnston {et~al.}(1993)Johnston, Lorimer, Harrison, Bailes, Lyne,
  Bell, Kaspi, Manchester, D'Amico, Nicastro, \& Jin}]{jlh+93}
Johnston, S., Lorimer, D.~R., Harrison, P.~A., Bailes, M., Lyne, A.~G., Bell,
  J.~F., Kaspi, V.~M., Manchester, R.~N., D'Amico, N., Nicastro, L., \& Jin, S.
  1993, Nature, 361, 613

\bibitem[{Johnston {et~al.}(1992{\natexlab{a}})Johnston, Lyne, Manchester,
  Kniffen, D'Amico, Lim, \& Ashworth}]{jlm+92}
Johnston, S., Lyne, A.~G., Manchester, R.~N., Kniffen, D.~A., D'Amico, N., Lim,
  J., \& Ashworth, M. 1992{\natexlab{a}}, MNRAS, 255, 401

\bibitem[{Johnston {et~al.}(1992{\natexlab{b}})Johnston, Manchester, Lyne,
  Bailes, Kaspi, Qiao, \& D'Amico}]{jml+92}
Johnston, S., Manchester, R.~N., Lyne, A.~G., Bailes, M., Kaspi, V.~M., Qiao,
  G., \& D'Amico, N. 1992{\natexlab{b}}, ApJ, 387, L37

\bibitem[{Johnston {et~al.}(1994)Johnston, Manchester, Lyne, Nicastro, \&
  Spyromilio}]{jml+94}
Johnston, S., Manchester, R.~N., Lyne, A.~G., Nicastro, L., \& Spyromilio, J.
  1994, MNRAS, 268, 430

\bibitem[{Johnston {et~al.}(1999)Johnston, Manchester, McConnell, \&
  Campbell-Wilson}]{jmmc99}
Johnston, S., Manchester, R.~N., McConnell, D., \& Campbell-Wilson, D. 1999,
  MNRAS, 302, 277

\bibitem[{Kaspi {et~al.}(1994)Kaspi, Johnston, Bell, Manchester, Bailes,
  Bessell, Lyne, \& D'Amico}]{kjb+94}
Kaspi, V.~M., Johnston, S., Bell, J.~F., Manchester, R.~N., Bailes, M.,
  Bessell, M., Lyne, A.~G., \& D'Amico, N. 1994, ApJ, 423, L43

\bibitem[{Kaspi {et~al.}(1995)Kaspi, Tavani, Nagase, Hirayama, Hoshino, Aoki,
  Kawai, \& Arons}]{ktn+95}
Kaspi, V.~M., Tavani, M., Nagase, F., Hirayama, M., Hoshino, M., Aoki, T.,
  Kawai, N., \& Arons, J. 1995, ApJ, 453, 424

\bibitem[{{Keith} {et~al.}(2009){Keith}, {Eatough}, {Lyne}, {Kramer},
  {Possenti}, {Camilo}, \& {Manchester}}]{kel+09}
{Keith}, M.~J., {Eatough}, R.~P., {Lyne}, A.~G., {Kramer}, M., {Possenti}, A.,
  {Camilo}, F., \& {Manchester}, R.~N. 2009, MNRAS, 395, 837

\bibitem[{Komesaroff {et~al.}(1973)Komesaroff, Ables, Cooke, Hamilton, \&
  McCulloch}]{kac+73}
Komesaroff, M.~M., Ables, J.~G., Cooke, D.~J., Hamilton, P.~A., \& McCulloch,
  P.~M. 1973, Astrophys. Lett., 15, 169

\bibitem[{{Kramer} {et~al.}(2003){Kramer}, {Bell}, {Manchester}, {Lyne},
  {Camilo}, {Stairs}, {D'Amico}, {Kaspi}, {Hobbs}, {Morris}, {Crawford},
  {Possenti}, {Joshi}, {McLaughlin}, {Lorimer}, \& {Faulkner}}]{kbm+03}
{Kramer}, M., {Bell}, J.~F., {Manchester}, R.~N., {Lyne}, A.~G., {Camilo}, F.,
  {Stairs}, I.~H., {D'Amico}, N., {Kaspi}, V.~M., {Hobbs}, G., {Morris}, D.~J.,
  {Crawford}, F., {Possenti}, A., {Joshi}, B.~C., {McLaughlin}, M.~A.,
  {Lorimer}, D.~R., \& {Faulkner}, A.~J. 2003, MNRAS, 342, 1299

\bibitem[{{Kramer} {et~al.}(2006){Kramer}, {Stairs}, {Manchester},
  {McLaughlin}, {Lyne}, {Ferdman}, {Burgay}, {Lorimer}, {Possenti}, {D'Amico},
  {Sarkissian}, {Hobbs}, {Reynolds}, {Freire}, \& {Camilo}}]{ksm+06}
{Kramer}, M., {Stairs}, I.~H., {Manchester}, R.~N., {McLaughlin}, M.~A.,
  {Lyne}, A.~G., {Ferdman}, R.~D., {Burgay}, M., {Lorimer}, D.~R., {Possenti},
  A., {D'Amico}, N., {Sarkissian}, J.~M., {Hobbs}, G.~B., {Reynolds}, J.~E.,
  {Freire}, P.~C.~C., \& {Camilo}, F. 2006, Science, 314, 97

\bibitem[{{Lorimer} {et~al.}(2006){Lorimer}, {Faulkner}, {Lyne}, {Manchester},
  {Kramer}, {McLaughlin}, {Hobbs}, {Possenti}, {Stairs}, {Camilo}, {Burgay},
  {D'Amico}, {Corongiu}, \& {Crawford}}]{lfl+06}
{Lorimer}, D.~R., {Faulkner}, A.~J., {Lyne}, A.~G., {Manchester}, R.~N.,
  {Kramer}, M., {McLaughlin}, M.~A., {Hobbs}, G., {Possenti}, A., {Stairs},
  I.~H., {Camilo}, F., {Burgay}, M., {D'Amico}, N., {Corongiu}, A., \&
  {Crawford}, F. 2006, MNRAS, 372, 777

\bibitem[{Lyne {et~al.}(1987)Lyne, Brinklow, Middleditch, Kulkarni, Backer, \&
  Clifton}]{lbm+87}
Lyne, A.~G., Brinklow, A., Middleditch, J., Kulkarni, S.~R., Backer, D.~C., \&
  Clifton, T.~R. 1987, Nature, 328, 399

\bibitem[{Lyne {et~al.}(2004)Lyne, Burgay, Kramer, Possenti, Manchester,
  Camilo, McLaughlin, Lorimer, D'Amico, Joshi, Reynolds, \& Freire}]{lbk+04}
Lyne, A.~G., Burgay, M., Kramer, M., Possenti, A., Manchester, R.~N., Camilo,
  F., McLaughlin, M.~A., Lorimer, D.~R., D'Amico, N., Joshi, B.~C., Reynolds,
  J., \& Freire, P. C.~C. 2004, Science, 303, 1153

\bibitem[{Lyne \& Manchester(1988)}]{lm88}
Lyne, A.~G. \& Manchester, R.~N. 1988, MNRAS, 234, 477

\bibitem[{Lyne {et~al.}(1998)Lyne, Manchester, Lorimer, Bailes, D'Amico,
  Tauris, Johnston, Bell, \& Nicastro}]{lml+98}
Lyne, A.~G., Manchester, R.~N., Lorimer, D.~R., Bailes, M., D'Amico, N.,
  Tauris, T.~M., Johnston, S., Bell, J.~F., \& Nicastro, L. 1998, MNRAS, 295,
  743

\bibitem[{{Lyutikov} \& {Thompson}(2005)}]{lt05}
{Lyutikov}, M. \& {Thompson}, C. 2005, ApJ, 634, 1223

\bibitem[{{Manchester}(1969)}]{man69}
{Manchester}, B.~A. 1969, Australian Journal of Physics Astrophysical
  Supplement, 12, 3

\bibitem[{{Manchester} \& {Goss}(1969)}]{mg69}
{Manchester}, B.~A. \& {Goss}, W.~M. 1969, Australian Journal of Physics
  Astrophysical Supplement, 11, 35

\bibitem[{Manchester(1971)}]{man71b}
Manchester, R.~N. 1971, ApJS, 23, 283

\bibitem[{{Manchester} {et~al.}(2006){Manchester}, {Fan}, {Lyne}, {Kaspi}, \&
  {Crawford}}]{mfl+06}
{Manchester}, R.~N., {Fan}, G., {Lyne}, A.~G., {Kaspi}, V.~M., \& {Crawford},
  F. 2006, ApJ, 649, 235

\bibitem[{Manchester {et~al.}(1980)Manchester, Hamilton, \& McCulloch}]{mhm80}
Manchester, R.~N., Hamilton, P.~A., \& McCulloch, P.~M. 1980, MNRAS, 192, 153

\bibitem[{{Manchester} {et~al.}(2005){Manchester}, {Hobbs}, {Teoh}, \&
  {Hobbs}}]{mhth05}
{Manchester}, R.~N., {Hobbs}, G.~B., {Teoh}, A., \& {Hobbs}, M. 2005, AJ, 129,
  1993

\bibitem[{Manchester {et~al.}(2001)Manchester, Lyne, Camilo, Bell, Kaspi,
  D'Amico, McKay, Crawford, Stairs, Possenti, Morris, \& Sheppard}]{mlc+01}
Manchester, R.~N., Lyne, A.~G., Camilo, F., Bell, J.~F., Kaspi, V.~M., D'Amico,
  N., McKay, N. P.~F., Crawford, F., Stairs, I.~H., Possenti, A., Morris,
  D.~J., \& Sheppard, D.~C. 2001, MNRAS, 328, 17

\bibitem[{Manchester {et~al.}(1996)Manchester, Lyne, D'Amico, Bailes, Johnston,
  Lorimer, Harrison, Nicastro, \& Bell}]{mld+96}
Manchester, R.~N., Lyne, A.~G., D'Amico, N., Bailes, M., Johnston, S., Lorimer,
  D.~R., Harrison, P.~A., Nicastro, L., \& Bell, J.~F. 1996, MNRAS, 279, 1235

\bibitem[{Manchester {et~al.}(1990)Manchester, Lyne, D'Amico, Johnston, Lim, \&
  Kniffen}]{mld+90}
Manchester, R.~N., Lyne, A.~G., D'Amico, N., Johnston, S., Lim, J., \& Kniffen,
  D.~A. 1990, Nature, 345, 598

\bibitem[{Manchester {et~al.}(1991)Manchester, Lyne, Robinson, D'Amico, Bailes,
  \& Lim}]{mlr+91}
Manchester, R.~N., Lyne, A.~G., Robinson, C., D'Amico, N., Bailes, M., \& Lim,
  J. 1991, Nature, 352, 219

\bibitem[{Manchester {et~al.}(1978)Manchester, Lyne, Taylor, Durdin, Large, \&
  Little}]{mlt+78}
Manchester, R.~N., Lyne, A.~G., Taylor, J.~H., Durdin, J.~M., Large, M.~I., \&
  Little, A.~G. 1978, MNRAS, 185, 409

\bibitem[{Manchester {et~al.}(1975)Manchester, Taylor, \& Huguenin}]{mth75}
Manchester, R.~N., Taylor, J.~H., \& Huguenin, G.~R. 1975, ApJ, 196, 83

\bibitem[{Marshall {et~al.}(1998)Marshall, Gotthelf, Zhang, Middleditch, \&
  Wang}]{mgz+98}
Marshall, F.~E., Gotthelf, E.~V., Zhang, W., Middleditch, J., \& Wang, Q.~D.
  1998, ApJ, 499, L179

\bibitem[{McConnell {et~al.}(1991)McConnell, McCulloch, Hamilton, Ables, Hall,
  Jacka, \& Hunt}]{mmh+91}
McConnell, D., McCulloch, P.~M., Hamilton, P.~A., Ables, J.~G., Hall, P.~J.,
  Jacka, C.~E., \& Hunt, A.~J. 1991, MNRAS, 249, 654

\bibitem[{McCulloch {et~al.}(1983)McCulloch, Hamilton, Ables, \& Hunt}]{mhah83}
McCulloch, P.~M., Hamilton, P.~A., Ables, J.~G., \& Hunt, A.~J. 1983, Nature,
  303, 307

\bibitem[{{McCulloch} {et~al.}(1972){McCulloch}, {Hamilton}, {Ables}, \&
  {Komesaroff}}]{mhak72}
{McCulloch}, P.~M., {Hamilton}, P.~A., {Ables}, J.~G., \& {Komesaroff}, M.~M.
  1972, Astrohys. Lett., 10, 163

\bibitem[{McCulloch {et~al.}(1976)McCulloch, Hamilton, Ables, \&
  Komesaroff}]{mhak76}
McCulloch, P.~M., Hamilton, P.~A., Ables, J.~G., \& Komesaroff, M.~M. 1976,
  MNRAS, 175, 71P

\bibitem[{McCulloch {et~al.}(1978)McCulloch, Hamilton, Manchester, \&
  Ables}]{mhma78}
McCulloch, P.~M., Hamilton, P.~A., Manchester, R.~N., \& Ables, J.~G. 1978,
  MNRAS, 183, 645

\bibitem[{{McLaughlin} {et~al.}(2004{\natexlab{a}}){McLaughlin}, {Kramer},
  {Lyne}, {Lorimer}, {Stairs}, {Possenti}, {Manchester}, {Freire}, {Joshi},
  {Burgay}, {Camilo}, \& {D'Amico}}]{mkl+04}
{McLaughlin}, M.~A., {Kramer}, M., {Lyne}, A.~G., {Lorimer}, D.~R., {Stairs},
  I.~H., {Possenti}, A., {Manchester}, R.~N., {Freire}, P.~C.~C., {Joshi},
  B.~C., {Burgay}, M., {Camilo}, F., \& {D'Amico}, N. 2004{\natexlab{a}}, ApJ,
  613, L57

\bibitem[{{McLaughlin} {et~al.}(2004{\natexlab{b}}){McLaughlin}, {Lyne},
  {Lorimer}, {Possenti}, {Manchester}, {Camilo}, {Stairs}, {Kramer}, {Burgay},
  {D'Amico}, {Freire}, {Joshi}, \& {Bhat}}]{mll+04}
{McLaughlin}, M.~A., {Lyne}, A.~G., {Lorimer}, D.~R., {Possenti}, A.,
  {Manchester}, R.~N., {Camilo}, F., {Stairs}, I.~H., {Kramer}, M., {Burgay},
  M., {D'Amico}, N., {Freire}, P.~C.~C., {Joshi}, B.~C., \& {Bhat}, N.~D.~R.
  2004{\natexlab{b}}, ApJ, 616, L131

\bibitem[{McLaughlin {et~al.}(2003)McLaughlin, Stairs, Kaspi, Lorimer, Kramer,
  Lyne, Manchester, Camilo, Hobbs, Possenti, D'Amico, \& Faulkner}]{msk+03}
McLaughlin, M.~A., Stairs, I.~H., Kaspi, V.~M., Lorimer, D.~R., Kramer, M.,
  Lyne, A.~G., Manchester, R.~N., Camilo, F., Hobbs, G., Possenti, A., D'Amico,
  N., \& Faulkner, A.~J. 2003, ApJ, 591, L135

\bibitem[{Morris {et~al.}(1981)Morris, Graham, Seiber, Bartel, \&
  Thomasson}]{mgs+81}
Morris, D., Graham, D.~A., Seiber, W., Bartel, N., \& Thomasson, P. 1981,
  A\&AS, 46, 421

\bibitem[{{Morris} {et~al.}(2002){Morris}, {Hobbs}, {Lyne}, {Stairs}, {Camilo},
  {Manchester}, {Possenti}, {Bell}, {Kaspi}, {Amico}, {McKay}, {Crawford}, \&
  {Kramer}}]{mhl+02}
{Morris}, D.~J., {Hobbs}, G., {Lyne}, A.~G., {Stairs}, I.~H., {Camilo}, F.,
  {Manchester}, R.~N., {Possenti}, A., {Bell}, J.~F., {Kaspi}, V.~M., {Amico},
  N.~D., {McKay}, N.~P.~F., {Crawford}, F., \& {Kramer}, M. 2002, MNRAS, 335,
  275

\bibitem[{{Perera} {et~al.}(2010){Perera}, {McLaughlin}, {Kramer}, {Stairs},
  {Ferdman}, {Freire}, {Possenti}, {Breton}, {Manchester}, {Burgay}, {Lyne}, \&
  {Camilo}}]{pmk+10}
{Perera}, B.~B.~P., {McLaughlin}, M.~A., {Kramer}, M., {Stairs}, I.~H.,
  {Ferdman}, R.~D., {Freire}, P.~C.~C., {Possenti}, A., {Breton}, R.~P.,
  {Manchester}, R.~N., {Burgay}, M., {Lyne}, A.~G., \& {Camilo}, F. 2010, ApJ,
  721, 1193

\bibitem[{{Possenti} {et~al.}(2003){Possenti}, {D'Amico}, {Manchester},
  {Camilo}, {Lyne}, {Sarkissian}, \& {Corongiu}}]{pdm+03}
{Possenti}, A., {D'Amico}, N., {Manchester}, R.~N., {Camilo}, F., {Lyne},
  A.~G., {Sarkissian}, J., \& {Corongiu}, A. 2003, ApJ, 599, 475

\bibitem[{Qiao {et~al.}(1995)Qiao, Manchester, Lyne, \& Gould}]{qmlg95}
Qiao, G.~J., Manchester, R.~N., Lyne, A.~G., \& Gould, D.~M. 1995, MNRAS, 274,
  572

\bibitem[{Radhakrishnan \& Cooke(1969)}]{rc69a}
Radhakrishnan, V. \& Cooke, D.~J. 1969, Astrophys. Lett., 3, 225

\bibitem[{Radhakrishnan {et~al.}(1969)Radhakrishnan, Cooke, Komesaroff, \&
  Morris}]{rckm69}
Radhakrishnan, V., Cooke, D.~J., Komesaroff, M.~M., \& Morris, D. 1969, Nature,
  221, 443

\bibitem[{Radhakrishnan {et~al.}(1968)Radhakrishnan, Komesaroff, \&
  Cooke}]{rkc68}
Radhakrishnan, V., Komesaroff, M.~M., \& Cooke, D.~J. 1968, Nature, 218, 229

\bibitem[{Radhakrishnan \& Manchester(1969)}]{rm69}
Radhakrishnan, V. \& Manchester, R.~N. 1969, Nature, 222, 228

\bibitem[{Rankin(1983)}]{ran83}
Rankin, J.~M. 1983, ApJ, 274, 333

\bibitem[{{Ransom} {et~al.}(2005){Ransom}, {Hessels}, {Stairs}, {Freire},
  {Camilo}, {Kaspi}, \& {Kaplan}}]{rhs+05}
{Ransom}, S.~M., {Hessels}, J.~W.~T., {Stairs}, I.~H., {Freire}, P.~C.~C.,
  {Camilo}, F., {Kaspi}, V.~M., \& {Kaplan}, D.~L. 2005, Science, 307, 892

\bibitem[{Robinson {et~al.}(1968)Robinson, Cooper, Gardiner, Wielebinski, \&
  Landecker}]{rcg+68}
Robinson, B.~J., Cooper, B. F.~C., Gardiner, F.~F., Wielebinski, R., \&
  Landecker, T.~L. 1968, Nature, 218, 1143

\bibitem[{Robinson {et~al.}(1995)Robinson, Lyne, Manchester, Bailes, D'Amico,
  \& Johnston}]{rlm+95}
Robinson, C.~R., Lyne, A.~G., Manchester, A.~G., Bailes, M., D'Amico, N., \&
  Johnston, S. 1995, MNRAS, 274, 547

\bibitem[{Seward {et~al.}(1984)Seward, Harnden, \& Helfand}]{shh84}
Seward, F.~D., Harnden, F.~R., \& Helfand, D.~J. 1984, ApJ, 287, L19

\bibitem[{Staveley-Smith {et~al.}(1996)Staveley-Smith, Wilson, Bird, Disney,
  Ekers, Freeman, Haynes, Sinclair, Vaile, Webster, \& Wright}]{swb+96}
Staveley-Smith, L., Wilson, W.~E., Bird, T.~S., Disney, M.~J., Ekers, R.~D.,
  Freeman, K.~C., Haynes, R.~F., Sinclair, M.~W., Vaile, R.~A., Webster, R.~L.,
  \& Wright, A.~E. 1996, PASA, 13, 243

\bibitem[{Stinebring {et~al.}(1984)Stinebring, Cordes, Rankin, Weisberg, \&
  Boriakoff}]{scr+84}
Stinebring, D.~R., Cordes, J.~M., Rankin, J.~M., Weisberg, J.~M., \& Boriakoff,
  V. 1984, ApJS, 55, 247

\bibitem[{van Ommen {et~al.}(1997)van Ommen, D'Alesssandro, Hamilton, \&
  McCulloch}]{vdhm97}
van Ommen, T.~D., D'Alesssandro, F.~D., Hamilton, P.~A., \& McCulloch, P.~M.
  1997, MNRAS, 287, 307

\bibitem[{van Straten {et~al.}(2001)van Straten, Bailes, Britton, Kulkarni,
  Anderson, Manchester, \& Sarkissian}]{vbb+01}
van Straten, W., Bailes, M., Britton, M., Kulkarni, S.~R., Anderson, S.~B.,
  Manchester, R.~N., \& Sarkissian, J. 2001, Nature, 412, 158

\bibitem[{{Verbiest} {et~al.}(2008){Verbiest}, {Bailes}, {van Straten},
  {Hobbs}, {Edwards}, {Manchester}, {Bhat}, {Sarkissian}, {Jacoby}, \&
  {Kulkarni}}]{vbv+08}
{Verbiest}, J.~P.~W., {Bailes}, M., {van Straten}, W., {Hobbs}, G.~B.,
  {Edwards}, R.~T., {Manchester}, R.~N., {Bhat}, N.~D.~R., {Sarkissian}, J.~M.,
  {Jacoby}, B.~A., \& {Kulkarni}, S.~R. 2008, ApJ, 679, 675

\bibitem[{{Weisberg} \& {Taylor}(2005)}]{wt05}
{Weisberg}, J.~M. \& {Taylor}, J.~H. 2005, in {Binary Radio Pulsars}, ed.
  F.~Rasio \& I.~H. Stairs (San Francisco: Astronomical Society of the
  Pacific), 25--31

\bibitem[{{Yan} {et~al.}(2011){Yan}, {Manchester}, {van Straten}, {Reynolds},
  {Hobbs}, {Wang}, {Bailes}, {Bhat}, {Burke-Spolaor}, {Champion}, {Coles},
  {Hotan}, {Khoo}, {Oslowski}, {Sarkissian}, {Verbiest}, \& {Yardley}}]{ymv+11}
{Yan}, W.~M., {Manchester}, R.~N., {van Straten}, W., {Reynolds}, J.~E.,
  {Hobbs}, G., {Wang}, N., {Bailes}, M., {Bhat}, N.~D.~R., {Burke-Spolaor}, S.,
  {Champion}, D.~J., {Coles}, W.~A., {Hotan}, A.~W., {Khoo}, J., {Oslowski},
  S., {Sarkissian}, J.~M., {Verbiest}, J.~P.~W., \& {Yardley}, D.~R.~B. 2011,
  MNRAS, 414, 2087

\end{thebibliography}

\end{document}